\documentclass[fleqn,10pt]{wlscirep}
\usepackage[utf8]{inputenc}
\usepackage[T1]{fontenc}
\usepackage{lineno}
\usepackage{graphicx}

\title{A Root-Zone Soil Salinity Observatory for Coastal Southwest Bangladesh}

\author[1,*]{Showmitra Kumar Sarkar}
\author[1]{Mafrid Haydar}
\author[1]{Rhyme Rubayet Rudra}
\author[1]{Tanmoy Mazumder}
\author[1]{Md. Sadmin Nur}
\author[1]{Md. Shahriar Islam}
\author[1]{Shakib Mohammad Sany}
\author[1]{Tanzim Al Noor}
\author[1]{Shakil Ahmed}
\author[2]{Myisha Ahmad}
\author[2]{Annajmus Sakib}
\author[3,*]{Sai Ravela}

\affil[1]{Department of Urban and Regional Planning, Khulna University of Engineering and Technology, Khulna-9203, Bangladesh}
\affil[2]{Climate Change Programme, BRAC, Dhaka-1212, Bangladesh}
\affil[3]{Earth Signals and Systems Group, Earth, Atmospheric and Planetary Sciences, Massachusetts Institute of Technology, Cambridge MA 01239, USA}

\affil[*]{corresponding author(s): Showmitra Kumar Sarkar (mail4dhrubo@gmail.com), Sai Ravela (ravela@mit.edu)}

\begin{abstract}
The research assesses soil salinity in the southwest coastal region of Bangladesh, collecting a total of 162 topsoil samples between March 1 and March 9, 2024, and processing them following the standard operating procedure for soil electrical conductivity (soil/water, 1:5). Electrical conductivity (EC) measurements obtained using a HI-6321 advanced conductivity benchtop meter were analyzed and visualized using bubble density mapping and the Empirical Bayesian Kriging interpolation method. The findings indicate that soil salinity in the study area ranges from 0.05 to 9.09 mS/cm, with the highest levels observed near Debhata and Koyra. A gradient of increasing soil salinity is clearly evident from the northern to southern regions. This dataset provides a critical resource for soil salinity-related research in the region, offering valuable insights to support decision-makers in understanding and mitigating the impacts of soil salinity in Bangladesh's coastal areas.
\end{abstract}
\begin{document}

\flushbottom
\maketitle
\thispagestyle{empty}

\section{Introduction}
Soil salinization, the accumulation of mineral salts in soil and water, arises from natural hydrologic processes and human activity~\cite{morshed2016soil}. While natural events—such as floods and storm surges—contribute to salinization, the primary drivers are often human-induced factors, including the overuse of fertilizers, inadequate drainage, poor irrigation practices, and unsustainable agricultural management~\cite{wu2008remote}. Globally, soil salinity poses a growing threat, affecting approximately one billion hectares of land, including 33 percent of irrigated agricultural fields and 20 percent of the world's cultivated lands~\cite{negacz2022saline}. This challenge is especially critical as global agricultural output must continue to meet the food demands of a rapidly increasing population.

According to the Food and Agriculture Organization (FAO), salinity now affects 424 million hectares of topsoil (0-30 cm) and 833 million hectares of subsoil (30-100 cm)—equivalent to about 3 percent of global topsoil and 6 percent of subsoil~\cite{FAO2024}. The economic implications are substantial: salinization results in an annual global loss of approximately USD 27 billion due to land degradation, reduced crop yields, and the abandonment of once-fertile fields. Alarmingly, the spread of salinity is accelerating, with up to two million hectares of land affected annually~\cite{morshed2021application}. Key drivers include insufficient precipitation, high surface evaporation in arid and semi-arid regions, saline water irrigation, and inadequate agricultural practices~\cite{nachshon2018cropland}.

Projections indicate that by 2050, over 50 percent of the world's arable land could be impacted by salinization if current trends continue~\cite{butcher2016soil}. This rapid expansion stems from poor land management and climate change, which intensifies droughts, alters rainfall patterns, and contributes to rising sea levels—particularly affecting coastal areas. These factors amplify the salinization threat, making it a pressing global environmental and agricultural issue.

\subsection{Salinity Threats to Bangladesh}
Bangladesh, among the most vulnerable nations to climate change and environmental degradation~\cite{sarkar2021simulation}, is facing a critical challenge with soil salinization, particularly in its coastal regions~\cite{wan2023estimating}. Situated in the low-lying Bengal Delta, Bangladesh is especially susceptible to rainfall~\cite{saha2024a,saha2024b}, flooding\cite{qiu2023}, and saltwater intrusion, which exacerbate the salinity issue~\cite{rudra2023artificial}. Notably, over 20 percent of Bangladesh's total land area and more than 30 percent of its arable land lie along its coastline, making salinization a significant threat to national food security and economic stability~\cite{wan2023estimating}. Already, varying salinity levels~\cite{hasan2019soil} impact 53 percent of coastal soils.

Over the past few decades, soil salinity in coastal Bangladesh has steadily worsened. Between 1973 and 2009, the area impacted by soil salinity grew by 26.7 percent—from 833,450 hectares to 1,056,190 hectares~\cite{sarkar2023partial}. Projections suggest that this expansion will persist, with an anticipated annual increase of 146 square kilometers of salt-affected land~\cite{morshed2021application}. Key drivers of this trend include reduced freshwater flows from upstream rivers, unpredictable rainfall, tidal amplification, storm surges, and insufficiently managed coastal polder systems~\cite{akash2024assessment}. These factors make salinity one of Bangladesh's most pressing environmental concerns, severely affecting agriculture, water resources, and public health~\cite{sarkar2023coupling}. Given the reliance of Bangladesh's coastal regions on rice production, shrimp farming, and other agricultural activities, the socio-economic impacts of salinization are substantial.

\subsection{Salinization in Coastal Bangladesh} Coastal Bangladesh, a region home to millions and integral to the nation's economy, faces heightened vulnerability to salinity-related impacts. Climate change has accelerated global sea level rise, significantly contributing to saltwater intrusion in these areas. From 1901 to 2018, global sea levels rose by 15–25 cm, with an average increase of 2.3 mm per year since 1970. This rate nearly doubled to 4.62 mm per year between 2013 and 2022~\cite{fatoric2012vulnerability}. In Bangladesh, sea levels have risen at an average rate of 5 mm annually over the past three decades~\cite{uzzaman2014impact}, intensifying the risk to coastal ecosystems and communities.

Saltwater intrusion into coastal freshwater systems—surface water and groundwater—severely diminishes agricultural productivity. Salinization within crop root zones reduces soil fertility, inhibits plant growth, and ultimately lowers crop yields~\cite{zorb2019salinity,umamaheswari2009should}. This intrusion also contaminates drinking water sources, putting public health at risk~\cite{wongsirikajorn2023high}. As coastal populations rely on these water sources for drinking and irrigation, they face increased health risks from high salt intake, including hypertension and cardiovascular diseases~\cite{haldar2017coping}.

The ramifications of salinity extend to infrastructure, leading to ground subsidence and disrupting local ecosystems vital to fisheries and other economic activities~\cite{sahbeni2023challenges}. Thus, the salinization of coastal areas in Bangladesh jeopardizes not only agriculture but also the fishing, trade, and tourism sectors, with significant impacts on food security, water availability, and the overall quality of life. Projections indicate that by 2100, sea level rise could result in flooding of 12.34 percent to 18 percent of coastal zones, further intensifying salinization and displacing millions~\cite{wongsirikajorn2023high}.

\subsection{Southwest Bangladesh}
The southwestern region of Bangladesh, home to the Sundarbans—the world's largest mangrove forest—is particularly vulnerable to soil salinization. This area frequently experiences tropical cyclones, which bring storm surges that inundate vast areas with saline water. Between 1877 and 1995, Bangladesh experienced 154 cyclones, many of which led to storm surges~\cite{dasgupta2015climate}. Over ten tropical cyclones have struck the country in recent years alone, impacting approximately 3.45 million people~\cite{sarkar2024cyclone}. These surges introduce high salt levels into agricultural lands, often rendering them unproductive for years. Unlike gradual salinization from sea-level rise, episodic salinization caused by cyclones can make the soil unsuitable for cultivation for up to a decade, severely affecting local food production.

Currently, soil salinity affects about 1.056 million hectares—nearly two-thirds of Bangladesh's coastal area~\cite{sarkar2023partial}. The southwest faces pronounced difficulty, where over-extraction of groundwater for irrigation has caused groundwater levels to drop, allowing seawater to seep into coastal aquifers~\cite{bhuyan2023spatio}. This infiltration has led to the long-term contamination of surface water and groundwater with salt, exacerbating the region's salinity challenges. Contributing factors include reduced freshwater inflows due to upstream dams, brackish-water prawn farming, and inadequate management of sluice gates and polders~\cite{akash2024assessment}.

With climate change driving rising sea levels and intensifying cyclones, more than 20 million people in southwestern Bangladesh face increasing risks from excessive salt intake through food and water sources~\cite{haldar2017coping}. The compounded effects of salinity on agriculture, water security, and public health make this region one of the most vulnerable in the world. As climate change accelerates salinization and amplifies environmental challenges, the long-term viability of farming, fishing, and other livelihoods is increasingly at risk. Consequently, research on salinity impacts has gained significant attention in Bangladesh.

\subsection{A Salinity Observatory}
Research on soil salinity in Bangladesh's coastal areas is expanding as various methods and approaches aim to address this pressing issue. For instance, studies like those by Sarkar et al.~\cite{sarkar2023partial} have employed partial least regression to analyze soil salinity. In contrast, Sarkar et al.~\cite{sarkar2023coupling} utilized machine learning techniques combined with satellite-based indices for improved detection accuracy. Other approaches have focused on remote sensing and spatial analyses, such as Morshed et al., who used satellite imagery to detect salinity levels across the region~\cite{morshed2016soil}. Meanwhile, Rezoyana et al. examined the impacts of salinity on coastal environments~\cite{rezoyana2023impact}, with Morshed et al. applying salinity-based zoning to guide land-use decisions~\cite{morshed2021application}. Studies by Kumar et al.~\cite{kumar2019seasonal} further highlighted relationships between soil salinity and other soil properties, adding valuable insights into salinity dynamics.

In specific regions like Khulna, local research has provided crucial salinity profiles, such as the work by Shaibur et al., who conducted an in-depth analysis of soil salinity gradients~\cite{shaibur2021gradients}. Fahim et al. explored how climate change influences salinity intrusion in coastal Bangladesh~\cite{fahim2024climate}. Other studies have examined broader impacts, including Hossain et al. and Rezoyana et al., who discussed salinity's effect on local livelihoods~\cite{hossain2010impact, rezoyana2023impact}. In the southwestern coastal region, Ashrafuzzaman et al.~\cite{ashrafuzzaman2022current} reviewed long-term trends of salinity intrusion. Additionally, researchers like Haldar et al. have investigated coping mechanisms for rice farming under saline conditions~\cite{haldar2017coping}. At the same time, Bhuyan et al. studied the spatio-temporal variability of soil and water salinity across the south-central coast~\cite{bhuyan2023spatio}. Akter et al.'s work on the hydrobiology of saline agricultural ecosystems~\cite{akter2023hydrobiology} underscores the environmental dimension of soil salinity.

However, field-based soil salinity measurements remain crucial for accurately assessing salinization's spatial extent and intensity, particularly in climate-vulnerable regions like the southwest coast. Field data on salinity provide vital, localized insights that satellite observations alone cannot capture, supporting effective adaptation measures in agriculture and ecosystem management. Data published by organizations like the Soil Resource Development Institute (SRDI)~\cite{SRDI2019}, the Department of Environment (DOE)~\cite{doe2023waterquality}, the Water Resources Planning Organization (WARPO), and the Bangladesh Water Development Board (BWDB)~\cite{BangladeshWaterDevelopmentBoard2022} play an essential role in establishing adaptive strategies. However, gaps persist; for example, SRDI's latest coastal region-based dataset is 14 years old~\cite{soil_salinity_report}, and few studies provide comprehensive, spatially dense data across both topsoil and river water in the southwest coastal belt.

Field-based soil salinity measurements across the southwest coastal region are crucial, particularly in climate change. These measurements help monitor and manage the impacts of rising sea levels and the increasing frequency of extreme weather events, both exacerbating soil salinity through saltwater intrusion and storm surges. By understanding salinity patterns in this region, researchers and policymakers can develop adaptive strategies to mitigate the detrimental effects on agricultural systems and local ecosystems. Accurate salinity data are essential for promoting sustainable land management practices and maintaining soil health and agricultural productivity in changing climate conditions.

Furthermore, these measurements are vital for informing climate-adaptive policies and regional planning and supporting sustainable development and environmental protection efforts in Bangladesh's coastal areas. Several organizations play a crucial role in publishing salinity data from field-based measurements, including the Soil Resource Development Institute (SRDI)~\cite{SRDI2019}, the Department of Environment (DOE)~\cite{doe2023waterquality}, the Water Resources Planning Organization (WARPO), and the Bangladesh Water Development Board (BWDB)~\cite{BangladeshWaterDevelopmentBoard2022}.

However, there are notable gaps in the available data. For example, although SRDI has published recent salinity data across the country, there has been insufficient focus on the specific conditions of the coastal regions~\cite{SRDI2022}. Additionally, the dataset lacks detail on how salinity characteristics vary across the southern coastal belt. The last coastal region-based dataset was published over 14 years ago by SRDI~\cite{soil_salinity_report}. Data published by the DOE~\cite{SRDI2019} and BWDB~\cite{BangladeshWaterDevelopmentBoard2022} primarily focus on salinity levels in major rivers and lakes without considering topsoil salinity or regional-level variations. Field-based soil salinity measurements across the southwest coastal region have been infrequently studied and have sparse spatial sampling. To our knowledge, no comprehensive regional-level analysis that densely samples river water and topsoil is publicly available. Furthermore, advanced modeling and geospatial analyses of salinity patterns, incorporating field survey data, have not received adequate attention.

\subsection{Focus of Study}
This study addresses these gaps by conducting a comprehensive soil sampling campaign at unprecedented density across three districts—Satkhira, Khulna, and Jessore—and analyzing the samples in the laboratory. We collected the data under specific soil conditions and land use (open fields, fallow land, dry soil). The resulting dataset offers a crucial resource for soil salinity research in the area, providing valuable insights for decision-makers in Bangladesh's coastal regions to understand better and address the ongoing and future impacts of soil salinity. It is also the first step in our work to establish a persistent community-driven co-active soil observatory~\cite{Ravela2018,Ravela2023}, where measurements calibrate and refine soil salinity models, whose predictions and predicted uncertainties then target subsequent measurements for efficacy and informativeness~\cite{trautner2020}.

\section*{Methods}
\subsection*{Sample collection from field}
We collected soil data from three southwest coastal districts of Bangladesh, namely Khulna, Satkhira, and Jessore, excluding the Sundarbans (Figure 1(a)). Our field data collection team comprised faculty, scientists, students, and professionals from the Massachusetts Institute of Technology (MIT), Khulna University of Engineering and Technology (KUET), and BRAC. The team was divided into five groups to conduct the soil-salinity data-collection campaign from March 1-9, 2024, in the study area. Before the campaign, we reviewed previous research papers, maps, documents produced by government authorities, and satellite images as foundational materials. Based on reconnaissance surveys, we developed five tentative routes connecting major union centers (Figure 1(b)). The survey teams visited each major union center to identify suitable locations for collecting soil samples. One hundred sixty-two soil samples were collected along these routes and near the major centers (Figure 1(b)). Additionally, we utilized river transport to collect samples from the riverbanks to the centers of the polders.

Our objective was to integrate soil salinity data with remote sensing data. Selecting an optimal location under the open sky required careful consideration to ensure accurate readings. We prioritized areas free from obstructions, such as trees and buildings, ideally situated alongside dry, open, and fallow fields (Figure 2). After identifying a suitable site, we utilized Google My Maps to obtain coordinates and captured a geocoded photograph to accurately mark the location (Figure 3(d)). At each selected site, we meticulously gathered 5 to 10 soil samples from an approximately 30m x 30m area. To collect the samples, we first cleared any surface disturbances and then excavated the soil to a depth of 30 cm (Figure 3(a)). Each sample was carefully placed in a sample bag (Figure 3(c)), ensuring proper geocoding within the bag (Figure 3(d)). Subsequently, we mixed each collected sample to create a homogenized composite that accurately represented a particular location. Extensive photographic documentation captures various angles of the site and the collection procedure. After completing the sampling at one site, the team proceeded to the following designated location, repeating the meticulous process to ensure comprehensive data collection.

\subsection*{Data processing in laboratory}
We processed the samples at the ESSG-WECG laboratory established at KUET according to the standard operating procedure for soil electrical conductivity (soil/water, 1:5) as proposed by the Food and Agriculture Organization of the United Nations\cite{FAO2021}. The sample testing procedure began with signing in on the timesheet and donning gloves and lab attire. We meticulously followed each step: a single packet was selected and moved to a clean location on the testing table, ensuring isolation from other packets nearby (Figure 4(a)). We mixed the soil within the packet thoroughly to eliminate clumps using shaking and, if necessary, a spatula. We spread a portion of the well-mixed soil delicately onto a clean foil plate. We resealed the working packet, and the foil plate was placed in the oven for five to ten minutes to reduce moisture content (Figure 4(b)). Larger particles were sieved out using a 1.0 mm mesh (Figure 4(c)), and the filtered soil was mixed with deionized water in a clean beaker at a ratio of 1:5 (Figure 4(d)). After stirring the mixture for 10 minutes (Figure 4(e)), it was allowed to rest for 30 minutes (Figure 4(f and g)). The conductivity and salinity were measured using a HI-6321 advanced conductivity benchtop meter (Figure 4(h and i)). We recorded readings alongside the sample number and electrical conductivity (EC)(Figure 4(j)) and took photographs of the instrument reading and test setup. After each sample analysis, we cleaned the equipment thoroughly and sanitized the workbench, ensuring an organized workspace for subsequent tests (Figure 4(k)).

\subsection*{Data Visualization}
In Figure 5, bubble density maps show the distribution of soil salinity (EC in mS/cm) in the research area using specified range and classes (from 0.05 mS/cm to 9.09 mS/cm in five classes). Most of the points are in the range of 0.05 mS/cm - 0.72 mS/cm, mainly in the top side (North-West and North-East) of the research area (including sample points in Jessore, upper Khulna, and upper and western Satkhira). Soil salinity in the range of 0.73 mS/cm -1.71 mS/cm is distributed mainly in the west side of Khulna and the mid and lower parts of Satkhira. Areas near Debhata, Shyamnagar, Paikgacha, Batiaghata, Dacope,  and Dumuria have soil salinity in the range of 1.72 mS/cm to 3.14 mS/cm. The soil salinity levels range from 3.15 mS/cm to 5.82 mS/cm near Debhata, Kaliganj, Shyamnagar, Koyra, Dacope,  Rampal, Tala, Paikgaccha, and Dumuria. The highest level of soil salinity (5.83 mS/cm - 9.09 mS/cm) is present near Debhata and Koyra. 

There are many interpolation techniques to visualize the soil salinity throughout the research area, such as IDW (Inverse Distance Weighted), Kriging, Natural Neighbor, Spline and Trend, etc. In previous studies, numerous researchers have examined the performance of different interpolation techniques in various geoenvironmental scenarios. Most results indicate that geostatistical methods outperform deterministic methods. A geostatistical method called Kriging uses information from adjacent sampled data points to estimate values in places with unobserved data, which makes it easier to see test results throughout the research area. It does so by giving nearby sample points weights, therefore addressing spatial correlation and producing forecasts that include measures of uncertainty in addition to reducing estimation error. We use Empirical Bayesian kriging to interpolate the data, and it shows the most accurate result for this dataset among several types of Kriging.

Figure 6 illustrates the distribution and horizontal-vertical variation of soil salinity throughout the research area. The cross-section C-C' indicates that soil salinity is increasing from west to east. The cross-section has an increasing slope for values from the starting point C. After reaching the peak (four mS/cm), there is a downward slope of the soil salinity values, but it is less steep here. The two longitudinal sections, A-A' and B-B', also show an increase in soil salinity from the north to south. The A-A' section depicts the values from low to moderate, while the B-B' section depicts the values from low to high. Where cross-section C-C' and longitudinal section C-C' intersect, areas adjacent to this point show the highest soil salinity level. Figure 7 shows how the salinity of the soil is distributed across Khulna, Satkhira, and nearby places using contour lines. EC levels range from 0.5 to 6.5 (mS/cm) In places like Debhata and Koyra, where red and orange contour lines show higher salinity levels.

\section*{Data Records}
The processes—which included selecting sample sites, collecting data, calibrating sensors, analyzing data in the lab, interpolating data, and archiving data— are complete. The geographical location of sample sites, data collection, and soil salinity values (EC in mS/cm, constitute the metadata. This dataset contains the soil salinity data for 162 sample sites gathered between March 1 and 9, 2024. This article's soil salinity dataset for Bangladesh's southwest coastal regions is available at this link: https://doi.org/10.5281/zenodo.14560019 and the dataset is in the XLSX format.

\section*{Technical Validation}
Firstly, we validate our findings and data by comparing them to existing literature (Following Table). Our study's soil salinity distribution corresponds closely with prior studies' findings, both in values and variability between different sections. We used five Ordinary Kriging, Universal kriging, Simple Kriging, Disjunctive kriging, and Empirical Bayesian kriging using ArcGIS 10.5 to visualize the soil salinity across the research area (Figure 8 and Figure 5). To evaluate interpolation results, we used the leave-one-out cross-validation method. We calculate the root mean square error (RMSE) from the cross-validation outputs. Empirical Bayesian Kriging shows the most accurate result with an RMSE of 1.3 (Figure 7 and Figure 2). Finally, we visualize the soil salinity scenario in the research area using Empirical Bayesian kriging with 162 data.

 \begin{table}[htb!]
\begin{tabular}{|p{3cm}|p{3cm}|p{3cm}|p{6cm}|}
\hline
Reference & Study Area & Soil Salinity Values & Brief Discussion \\
\hline
(Morshed, Islam, \& Jamil, 2016) & Khulna, Patuakhali, Pirojpur, Barisal, Bagerhat, Noakhali, Lakshmipur, Chandpur, Bhola, and Gopalgonj &   Ranged from Less than 1 – Greater than 8 (EC in dS/m or mS/cm) & Soil Salinity was disturbed in 6 classes. Most areas are in the Low saline (EC, 2–4) and medium saline (EC, 4–6) classes, which comprise 31.17\% and 29.99\% of the area, respectively. The outputs are like this study.  \\
\hline 
(Morshed et al., 2021) & Jessore, Khulna, and Satkhira & Ranged from Less than 1 – Greater than 3 (EC in dS/m or mS/cm) & In 2016, the salinity levels in the study area were notably elevated, particularly in the southern region. Approximately 55.80\% of the area had low salinity, 25.44\% had medium salinity, and 18.76\% had high salinity. Furthermore, the spatial distribution and sectional variation of soil salinity align with our research findings. \\ 
\hline
(Sarkar et al., 2023) & Satkhira & - & The analysis reveals that the salinity levels classified as extremely high and high are around 29.02\% and 13.55\%, respectively. The upazilas of Debhata, Assasuni, and Shaymnagar exhibit elevated levels of salt, ranging from high to exceedingly high 5. The classes of soil salinity change from extremely low to extremely high for north-south and east-west movement, like this study.  \\
\hline
(Sarkar et al., 2023) & Satkhira & - & The total area affected by high soil salinity is documented as 977.94 square kilometers or approximately 43.51\% of the entire research area. Furthermore, 30.56\% of the Satkhira region is affected by moderate soil salinity, covering an area of 686.92 square kilometers. In comparison, low soil salinity affects 25.93\% of the land, totaling 582.73 square kilometers 16. The levels of soil salinity vary from low (0) to high (1) when moving from north to south and from east to west, as shown in this study. \\
\hline
\end{tabular}
\end{table}

\section*{Acknowledgment}
This research is part of the MIT Climate Grand Challenge Jameel Observatory CREWSNet and Weather and Climate Extremes projects. Schmidt Sciences, LLC and Liberty Mutual (029024-00020) also supported this work.


\bibliography{sample}

\newpage
\section*{Figures \& Tables}

\begin{figure}[h]
\centering
\includegraphics[width=\linewidth]{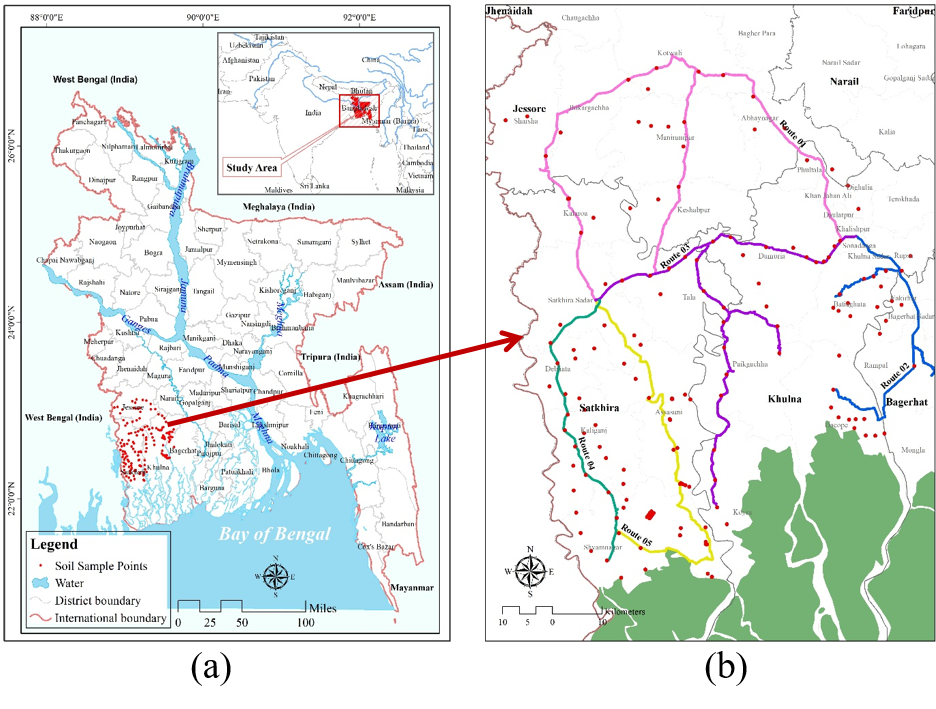}
\caption{(a) Location of study area in context of Bangladesh; (b) soil sample points and routes. Routes are color-coded for each of the five teams. }
\label{fig: Figure 1}
\end{figure}

\begin{figure}[h]
\centering
\includegraphics[width=\linewidth]{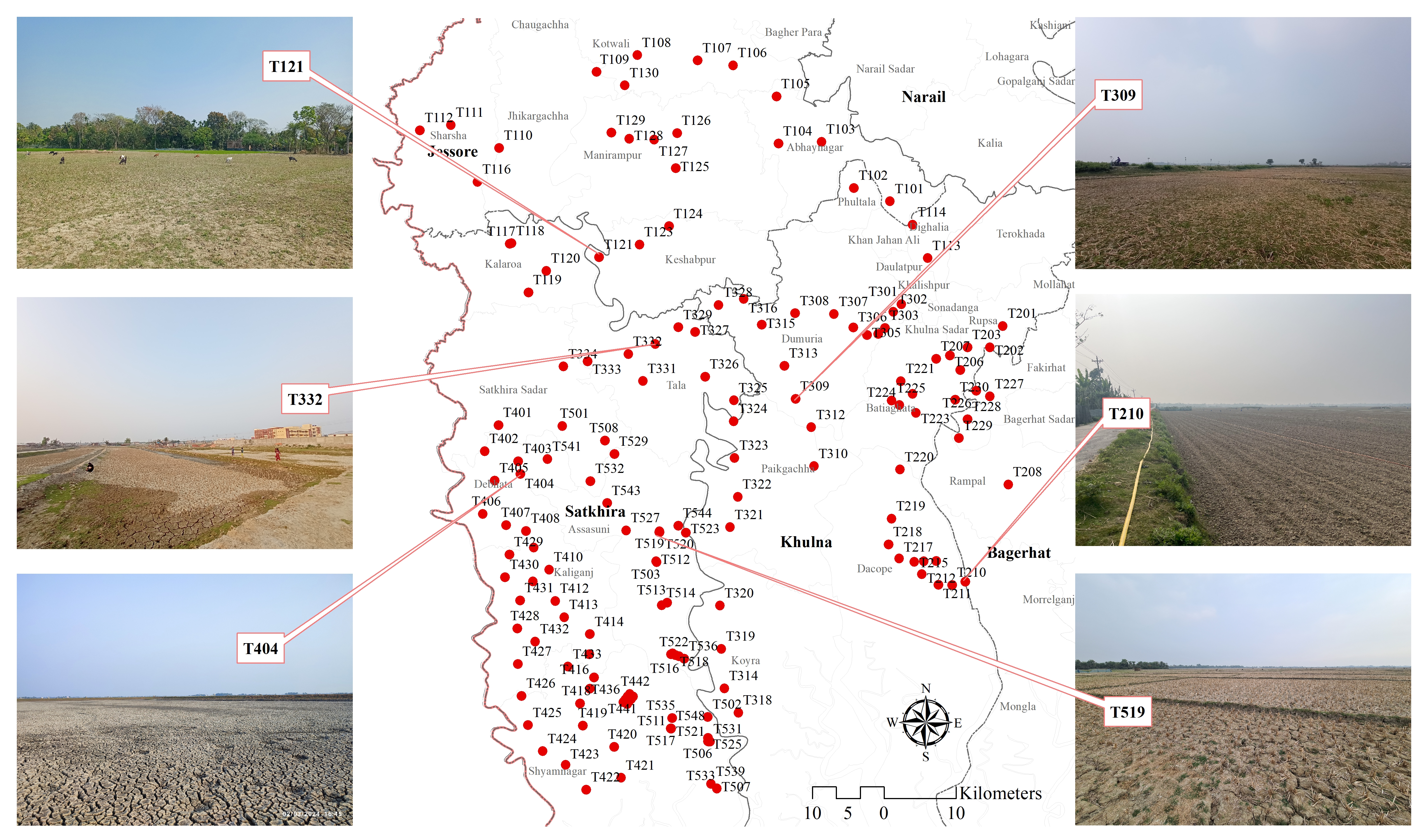}
\caption{Overview of soil sample sites with geocodes (tags) labeled for each site. We collected samples over open and fallow fields with dry soil along with river and water samples along various routes. }
\label{fig:Figure 2}
\end{figure}

\begin{figure}[h]
\centering
\includegraphics[width=\linewidth]{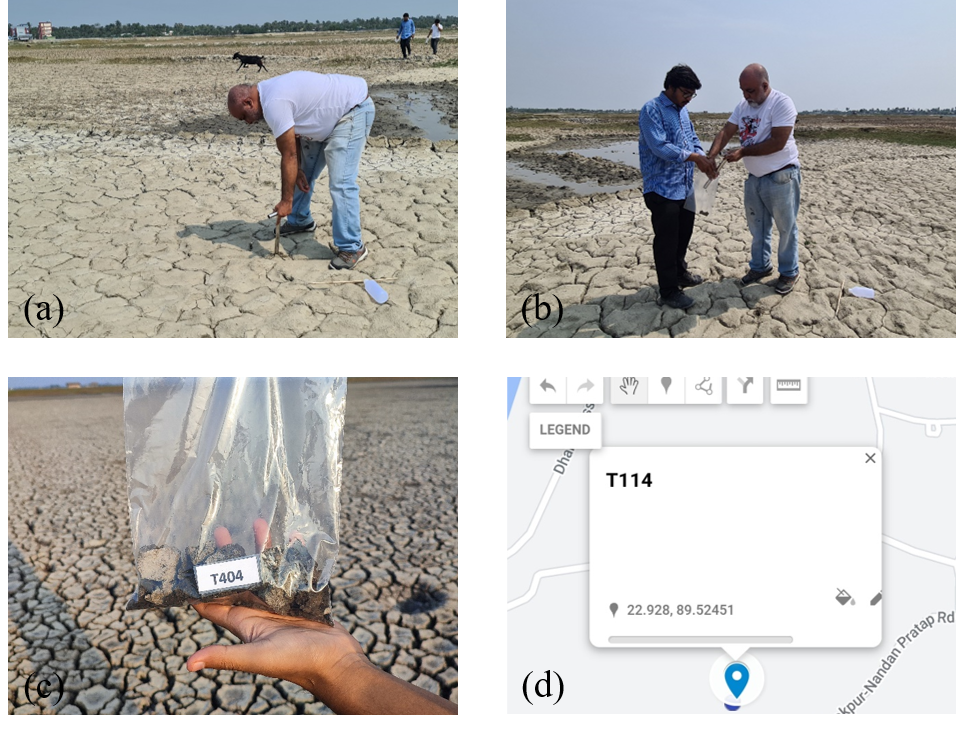}
\caption{(a) Collecting a soil sample from the ground; (b) placing the sample in a sample bag; (c) adding proper geocoding information to the bag; (d) obtaining coordinates using Google My Maps. We collected the soil over the top 30cm.}
\label{fig:Figure 3}
\end{figure}

\begin{figure}[h]
\centering
\includegraphics[width=\linewidth]{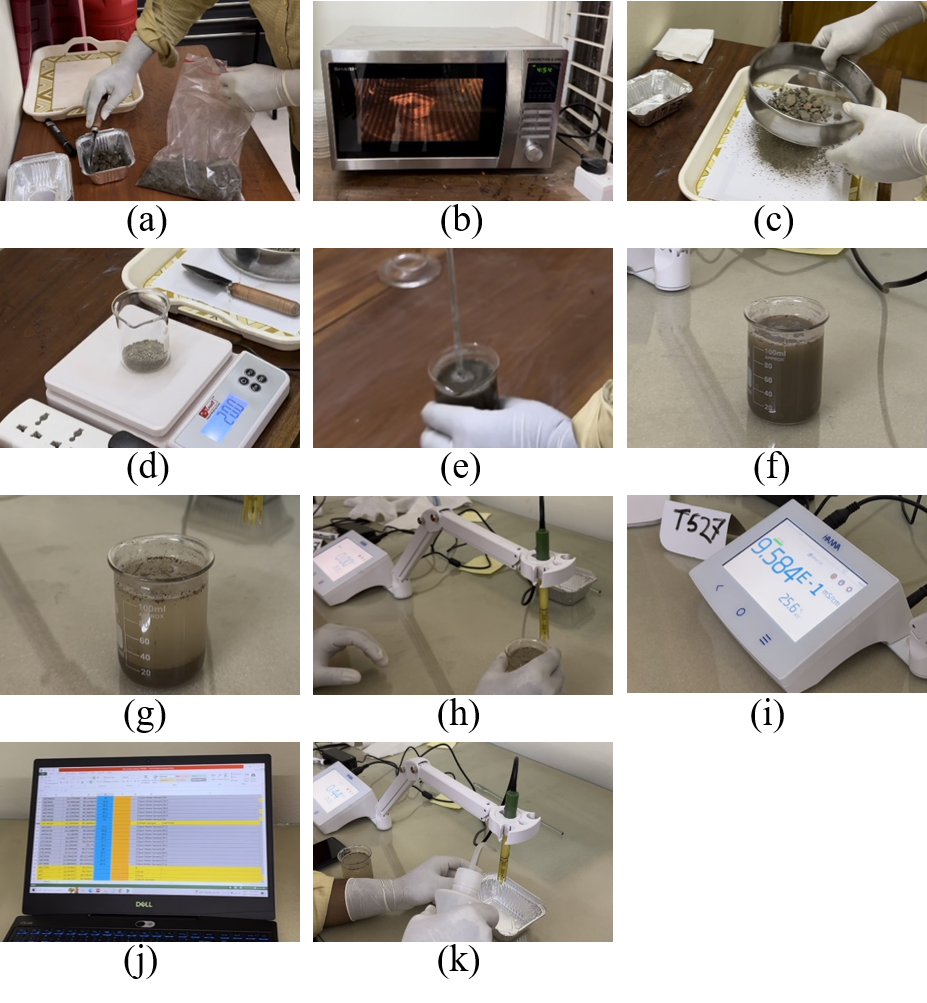}
\caption{(a) Collecting soil from sample bag; (b) drying the soil in oven; (c) sieving the soil; (d) weighing the soil; (e) stirring the mixture; (f) placing the mixture to rest; (g) settling down of the mixture; (h) testing the mixture; (i) taking photograph of data; (j) storing the data; (k) cleaning the equipment.}
\label{fig:Figure 4}
\end{figure}

\begin{figure}[h]
\centering
\includegraphics[width=0.8\linewidth]{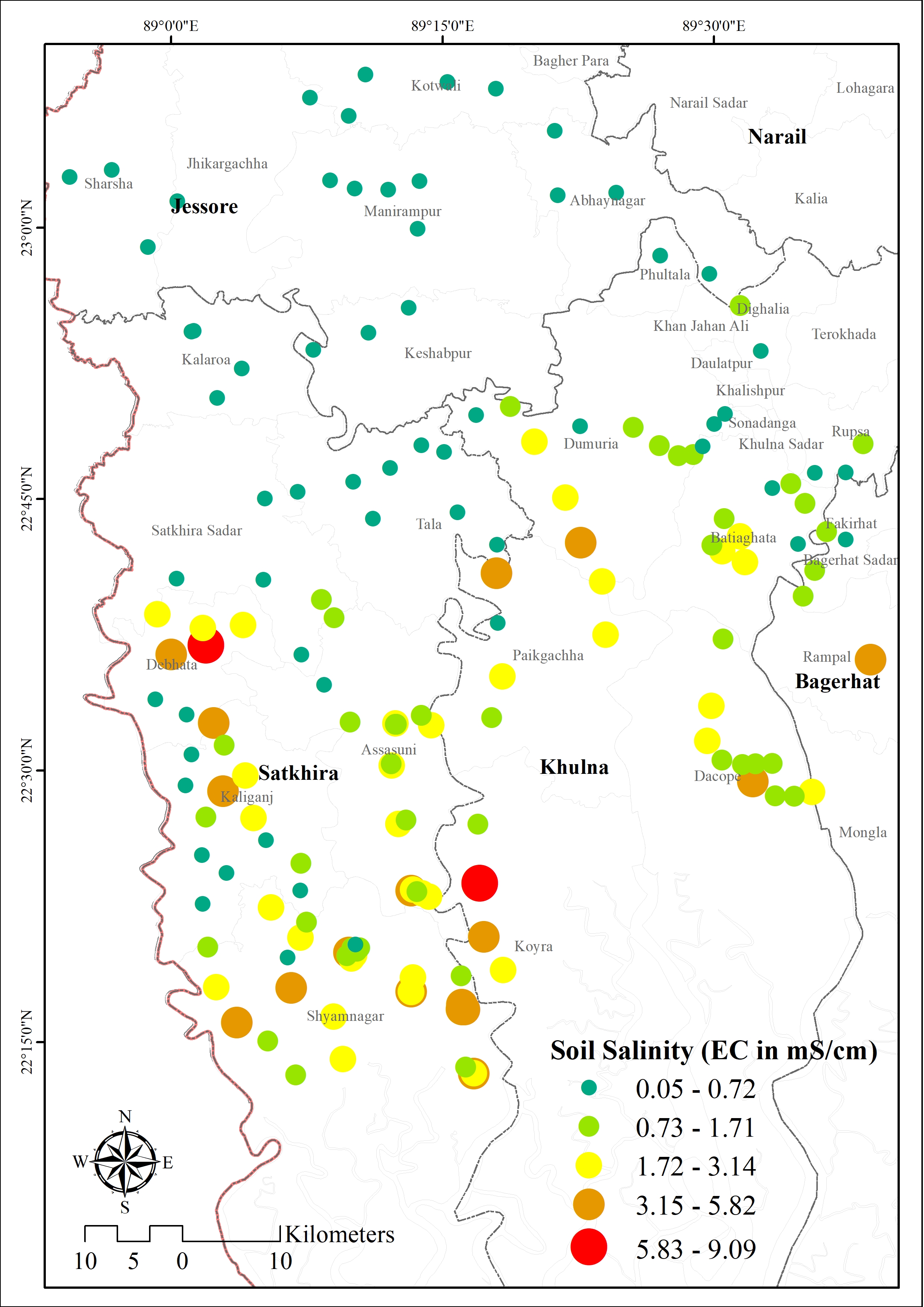}
\caption{Distribution of soil salinity in the study area colored and sized by electrical conductivity (EC) measure of salinity.}
\label{fig:Figure 5}
\end{figure}

\begin{figure}[ht]
\centering
\includegraphics[width=\linewidth]{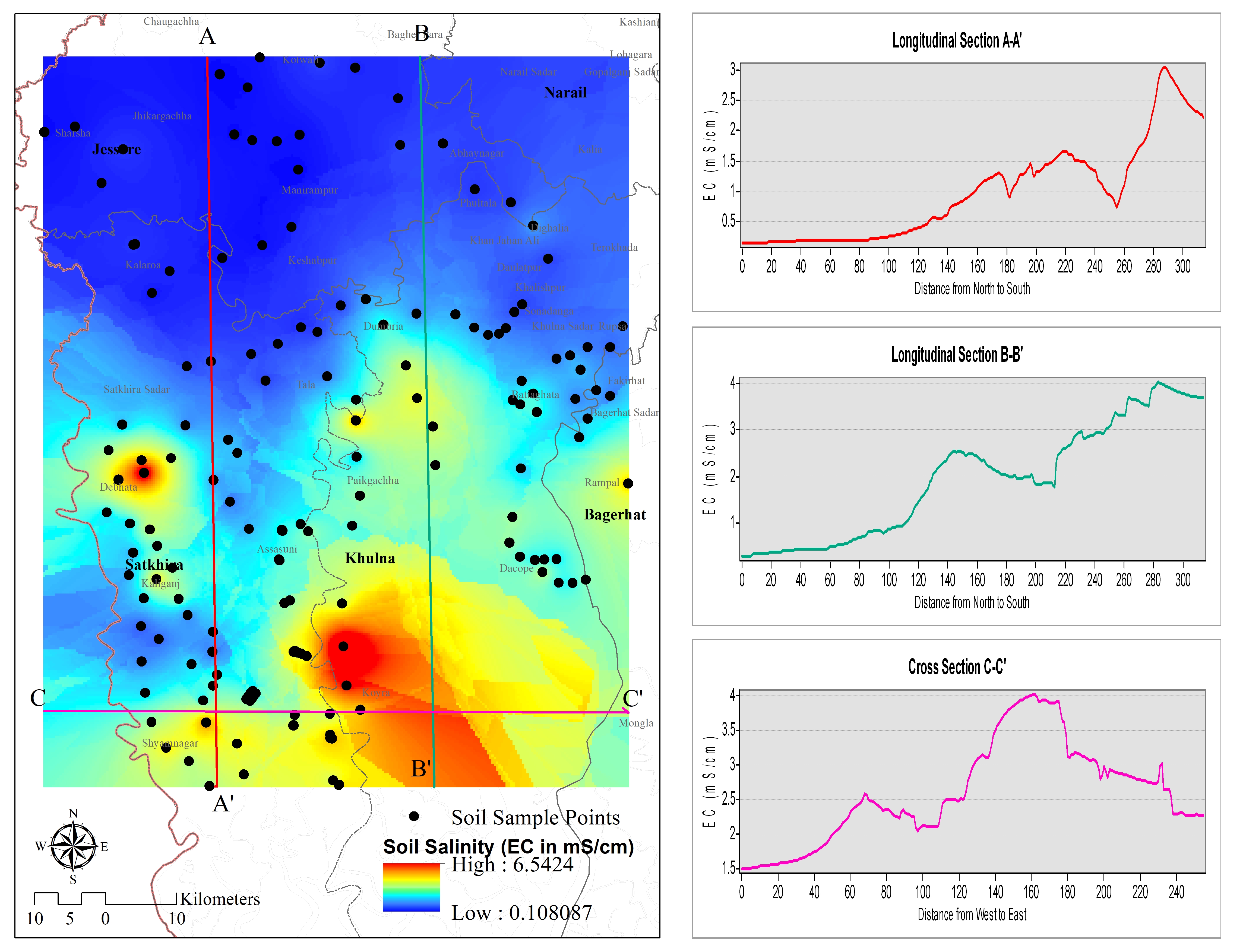}
\caption{Sectional variation of soil salinity in the study area. The graphs on the right plot salinity values along transects marked in the left plot.}
\label{fig:Figure 6}
\end{figure}

\begin{figure}[h]
\centering
\includegraphics[width=\linewidth]{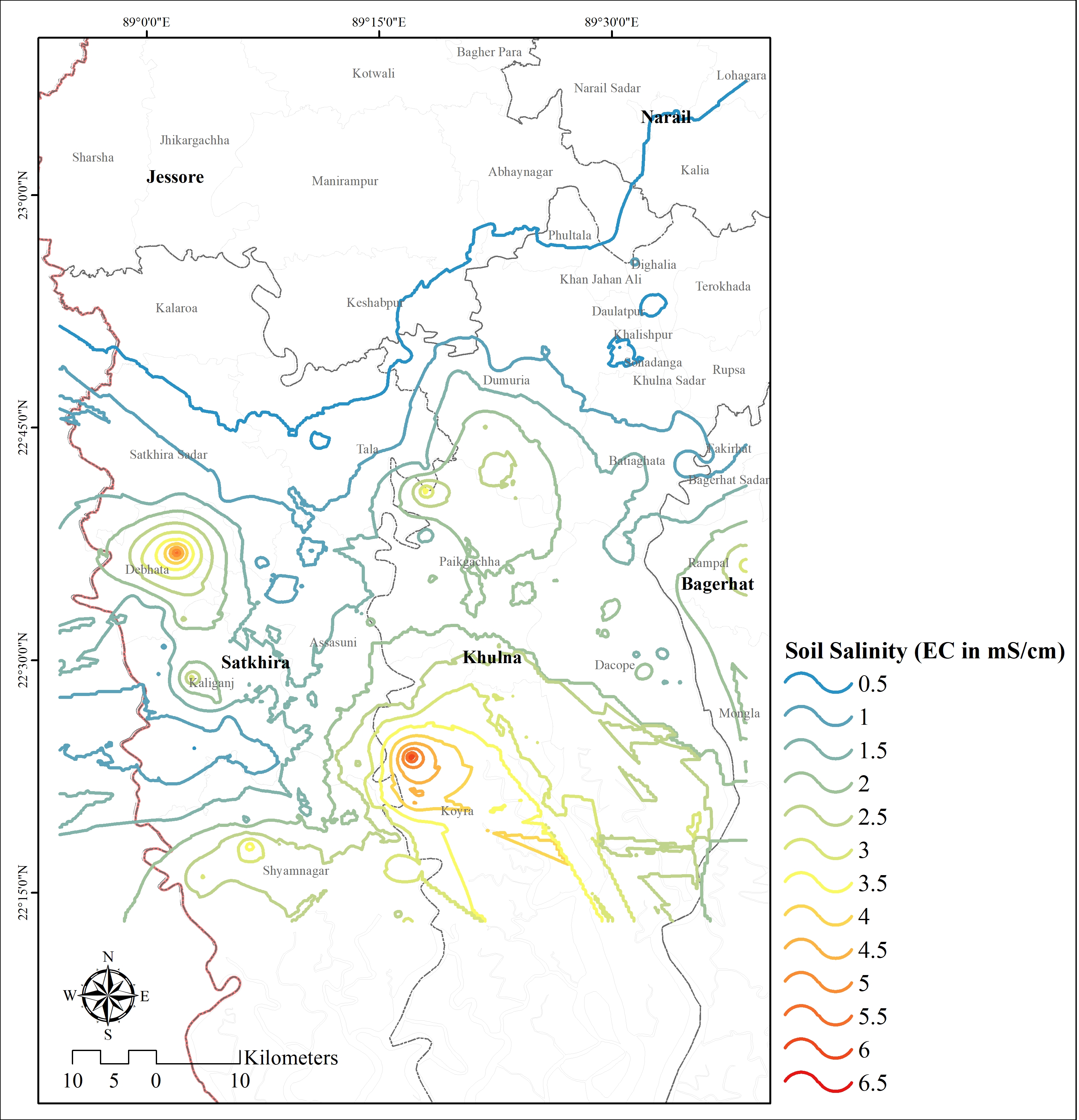}
\caption{Contour distribution of soil salinity in the study area}
\label{fig:Figure 7}
\end{figure}

\begin{figure}[h]
\centering
\includegraphics[width=0.7\linewidth]{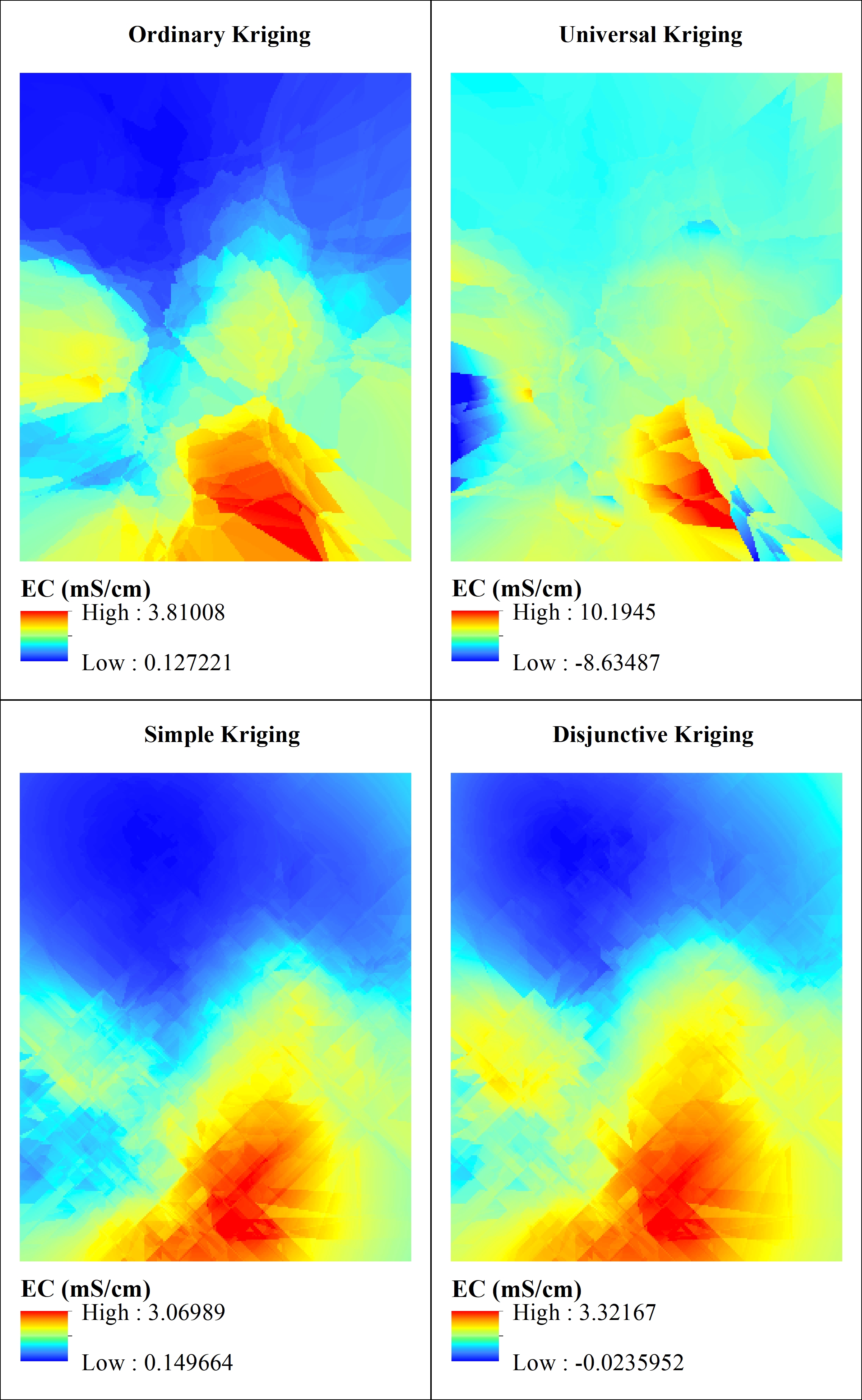}
\caption{ Interpolation of soil salinity using different types of Kriging. }
\label{fig: Figure 8}
\end{figure}

\begin{figure}[h]
\centering
\includegraphics[width=\linewidth]{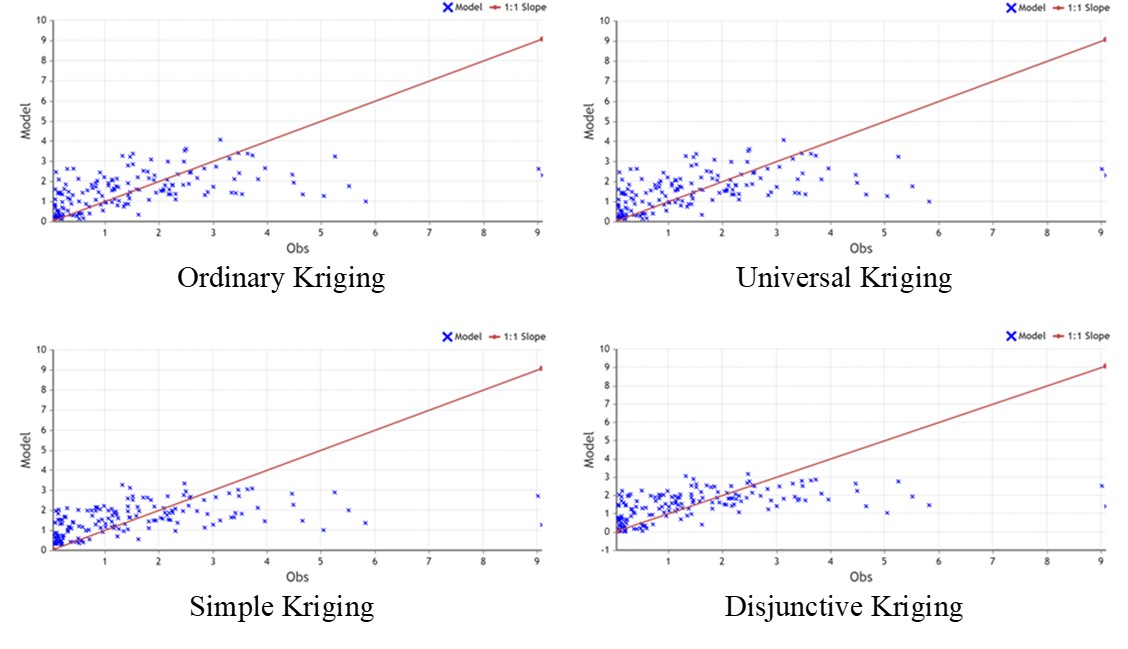}
\caption{Distribution of observed and model outputs from different types of Kriging at test locations.}
\label{fig: Figure 9}
\end{figure}

\end{document}